# A Quantum Genetic Algorithm-Enhanced Self-Supervised Intrusion Detection System for Wireless Sensor Networks in the Internet of Things

## Hamid Barati et al.

Department of Computer Engineering, Dezful Branch, Islamic Azad University, Dezful, Iran

**Abstract**

The rapid expansion of the Internet of Things (IoT) and Wireless Sensor Networks (WSNs) has significantly increased the attack surface of such systems, making them vulnerable to a wide range of cyber threats. Traditional Intrusion Detection Systems (IDS) often fail to meet the stringent requirements of resource-constrained IoT environments due to their high computational cost and reliance on large labeled datasets. To address these challenges, this paper proposes a novel hybrid Intrusion Detection System that integrates a Quantum Genetic Algorithm (QGA) with Self-Supervised Learning (SSL). The QGA leverages quantum-inspired evolutionary operators to optimize feature selection and fine-tune model parameters, ensuring lightweight yet efficient detection in resource-limited networks. Meanwhile, SSL enables the system to learn robust representations from unlabeled data, thereby reducing dependency on manually labeled training sets. The proposed framework is evaluated on benchmark IoT intrusion datasets, demonstrating superior performance in terms of detection accuracy, false positive rate, and computational efficiency compared to conventional evolutionary and deep learning-based IDS models. The results highlight the potential of combining quantum-inspired optimization with self-supervised paradigms to design next-generation intrusion detection solutions for IoT and WSN environments.

## 1. Introduction

The rapid proliferation of the Internet of Things (IoT) and Wireless Sensor Networks (WSNs) has revolutionized a wide spectrum of applications, ranging from smart healthcare and intelligent transportation systems to industrial automation and critical infrastructure monitoring. These networks consist of a large number of heterogeneous, resource-constrained devices that collect, process, and exchange sensitive information [1-4]. However, the openness of communication channels, limited computational power, and the distributed nature of IoT and WSNs make them highly vulnerable to diverse cyber threats, including denial-of-service (DoS), sinkhole, Sybil, and data injection attacks. Ensuring the security and reliability of these networks has therefore become a major challenge and a pressing research priority [5-7].
Intrusion Detection Systems (IDS) have emerged as an essential component in strengthening IoT and WSN security by monitoring network traffic and identifying malicious activities. Conventional IDS approaches can be broadly classified into signature-based and anomaly-based detection. While signature-based methods achieve high accuracy for known attacks, they fail to detect novel or zero-day threats. On the other hand, anomaly-based methods are more effective against unknown attacks but often suffer from high false positive rates and computational overhead, which are unsuitable for low-power devices in WSNs. Moreover, most existing IDS solutions rely heavily on large amounts of labeled training data, which is difficult to obtain in real-world IoT deployments [8-13].

To overcome these limitations, researchers have increasingly explored machine learning (ML) and optimization algorithms for building intelligent IDS models. Evolutionary algorithms such as Genetic Algorithm (GA), Particle Swarm Optimization (PSO), and Harris Hawks Optimization (HHO) have been widely used for feature selection and parameter tuning. However, classical evolutionary methods often converge slowly and may get trapped in local optima, limiting their effectiveness in highly dynamic IoT environments. Recent studies suggest that quantum-inspired algorithms, which exploit the principles of quantum superposition and entanglement, can provide stronger global search capabilities compared to their classical counterparts. In particular, the Quantum Genetic Algorithm (QGA) has demonstrated superior convergence speed and optimization performance in various domains, yet its potential in IoT security remains largely unexplored.

In parallel, advances in Self-Supervised Learning (SSL) have opened new opportunities for developing robust IDS solutions. Unlike traditional supervised learning, SSL can learn discriminative feature representations from unlabeled data by designing pretext tasks, significantly reducing the dependence on expensive labeled datasets. Given the abundance of unlabeled traffic data in IoT and WSNs, SSL provides a promising avenue for training efficient IDS models capable of detecting both known and unknown intrusions.

Motivated by these challenges and opportunities, this paper introduces a novel IDS framework that integrates Quantum Genetic Algorithm-based optimization with Self-Supervised Learning for intrusion detection in WSNs. The QGA is employed to optimize feature selection and model parameters, while SSL enables effective learning from unlabeled data, thereby reducing the reliance on manual annotation. The proposed system aims to achieve high detection accuracy, reduced false positive rates, and low computational cost, making it suitable for resource-constrained IoT environments.

The main contributions of this work are summarized as follows:
1. A novel hybrid IDS framework that combines Quantum Genetic Algorithm and Self-Supervised Learning for WSN-based IoT environments.
2. Efficient feature selection and parameter optimization using QGA to enhance detection accuracy while minimizing computational complexity.
3. Leveraging unlabeled IoT traffic data through SSL to reduce dependency on costly labeled datasets.
4. Comprehensive evaluation on benchmark IoT intrusion datasets, demonstrating superior performance compared to state-of-the-art IDS models.

The remainder of this paper is structured as follows. Section 2 reviews related work in IDS for IoT and WSNs. Section 3 presents the proposed methodology, including the integration of QGA and SSL. Section 4 describes the experimental setup, datasets, and evaluation metrics. Section 5 discusses the results and comparative analysis. Finally, Section 6 concludes the paper and outlines directions for future research.

## 2. Related Work

In recent years, researchers have devoted significant efforts to developing intelligent intrusion detection systems (IDS) for IoT and WSN environments. A variety of approaches ranging from self-supervised learning (SSL) and federated learning (FL) to evolutionary optimization and lightweight deep learning architectures have been proposed.

[14] Applying Self-Supervised Learning to Network Intrusion Detection (2024) introduced an SSL-based framework for traffic representation learning without reliance on labeled data. By exploiting graph neural networks and contrastive objectives, the system significantly reduced labeling costs while achieving improved accuracy on benchmark datasets, especially in imbalanced conditions .

[15] Contrastive Learning for Network Intrusion Detection (2025) provided a systematic review of contrastive learning methods for intrusion detection. It highlighted challenges such as the selection of positive/negative samples, data imbalance, and generalization to emerging attacks, and identified research gaps for future SSL-based IDS designs .

[16] In FS3 (Few-Shot and Self-Supervised Framework, 2023), the authors combined SSL with few-shot learning for IDS in IoT. The framework leveraged limited labeled data with abundant unlabeled samples, enabling effective intrusion detection in resource-limited IoT environments. The approach yielded higher F1-scores and lower false positive rates compared to conventional supervised models .

[17] End-to-End Intrusion Detection with Contrastive Learning (2024) proposed a hierarchical deep model based on CNNs and transformers trained via contrastive objectives. Without relying on handcrafted features, the system achieved superior performance across multiple datasets, demonstrating the scalability of SSL-based end-to-end architectures .

[18] Self-Supervised Transformer-based IDS (2025) leveraged transformer encoders trained with self-supervised pretext tasks on raw network flows. The model captured long-range dependencies and achieved competitive accuracy with limited labeled data, showing strong potential for anomaly detection in IoT networks .

[19] Enhancing IoT Security through SSL with MarkovGCN (2024) employed a Markov-based graph convolutional network for learning robust representations of IoT device behaviors. By incorporating SSL objectives, the method improved sensitivity to rare attacks and outperformed baseline supervised GCN-based IDS solutions .

[20] Contrastive Learning with Bayesian Gaussian Mixture Models (2025) presented a hybrid approach where SSL representations were clustered with BGMM for anomaly detection. This method was particularly effective in handling rare classes and imbalanced datasets, resulting in improved recall for minority attack types .

[21] Federated Learning-Based Intrusion Detection (2024) systematically evaluated IDS training across distributed IoT nodes using FL. The study investigated trade-offs between local data volume, model aggregation strategies, and communication cost, ultimately demonstrating the feasibility of FL for preserving privacy in IoT IDS .

[22] FL-based IDS for IoT (2024) proposed hybrid supervised/unsupervised models under federated settings. The study emphasized deployment feasibility in constrained IoT devices and showed that federated aggregation can maintain competitive accuracy while minimizing central data sharing .

[23] WOGRU-IDS (2022) optimized Gated Recurrent Units using the Whale Optimization Algorithm for IoT/WSN intrusion detection. The model improved classification accuracy and stability compared to baseline deep learning models, highlighting the potential of bio-inspired metaheuristics in IDS .

[24] Deep Learning-based IDS on UNSW-NB15 (2025) developed a robust pipeline using CNNs for feature extraction and classification. The system emphasized stable training on modern datasets, yielding better AUC and precision than prior baselines .

[25] DL-based Attack Detection for IoT (2025) focused on optimizing preprocessing pipelines and handling imbalanced data. Evaluations on NSL-KDD and CIC-IDS-2017 confirmed that the method achieved higher detection accuracy for IoT-specific attack scenarios .

[26] Lightweight IDS for IoT (2023) proposed a resource-efficient deep model tested on CIC-IDS2017, N-BaIoT, and CICIoT2023 datasets. It reduced computational complexity while maintaining high detection accuracy, making it suitable for deployment on edge devices .

[27] Rough-Fuzzy + Parallel Quantum Genetic Algorithm IDS (2023) integrated rough-fuzzy feature selection with a parallelized QGA optimization strategy. The approach accelerated convergence and improved detection performance, highlighting the role of quantum-inspired algorithms in IDS optimization .

[28] Intelligent Parameter-based IDS for IoT (2024) explored smart parameter tuning with machine learning classifiers on UNSW-NB15 and BoT-IoT. The study reported a balance between detection accuracy and computational cost, making it a viable candidate for real-time IoT deployment.

Comparative Summary of Related Works

| Ref | Year | Dataset(s) | Approach | Strengths | Limitations |
|---|---|---|---|---|---|
| [14] | 2024 | NSL-KDD, UNSW-NB15 | SSL + GNN | Reduced labeling cost, better in imbalance | High training overhead |
| [15] | 2024 | Multiple | Survey (Contrastive SSL) | Comprehensive analysis, identifies gaps | No experimental model |
| [16] | 2023 | IoT datasets | SSL + Few-Shot | Works with minimal labels | Complexity in multi-task training |
| [17] | 2024 | Multiple | End-to-end CNN+Transformer | High scalability, no handcrafted features | Model size large |
| [18] | 2025 | CIC-IDS, NSL-KDD | SSL + Transformer | Captures long dependencies | Needs pretraining resources |
| [19] | 2024 | IoT traffic | SSL + MarkovGCN | Sensitive to rare attacks | Graph modeling overhead |
| [20] | 2025 | UNSW-NB15 | SSL + BGMM | Effective on imbalanced data | High inference latency |
| [21] | 2025 | IoT nodes | Federated Learning | Privacy-preserving | Communication overhead |
| [22] | 2025 | IoT datasets | FL Hybrid Models | Deployment feasibility | Lower accuracy than centralized DL |
| [23] | 2022 | IoT/WSN | WOA + GRU | Better accuracy & stability | Metaheuristic tuning cost |
| [24] | 2025 | UNSW-NB15 | CNN-based IDS | High precision & AUC | Needs GPU acceleration |
| [25] | 2025 | NSL-KDD, CIC-IDS2017 | DL IDS | Improved preprocessing & balance | Dataset-specific |
| [26] | 2023 | CIC-IDS2017, N-BaIoT | Lightweight DL | Edge-friendly | Slightly lower accuracy |
| [27] | 2024 | NSL-KDD, UNSW | Rough-Fuzzy + QGA | Faster convergence, better features | Parallelism complexity |
| [28] | 2025 | UNSW-NB15, BoT-IoT | Smart ML Parameter IDS | Balance accuracy/cost | Limited to classical ML |

## 3. Proposed Methodology

This section introduces the proposed Quantum Genetic Algorithm-Enhanced Self-Supervised Intrusion Detection System (QGA-SSL IDS) for Wireless Sensor Networks (WSNs) in the Internet of Things (IoT). The methodology is designed to address three fundamental challenges: (i) the scarcity of labeled traffic data for intrusion detection, (ii) the need for lightweight yet accurate detection in resource-constrained WSN nodes, and (iii) the necessity of rapid adaptation to novel and dynamic attack behaviors. The proposed framework integrates self-supervised representation learning with quantum-inspired evolutionary optimization, yielding a scalable IDS that achieves high accuracy with minimal resource cost.

The overall workflow of the methodology is presented through six tightly coupled stages: data acquisition, preprocessing, SSL-based representation learning, QGA-driven feature and parameter optimization, intrusion classification, and adaptive response. Each stage contributes uniquely to the transformation of raw traffic into actionable intrusion detection, as detailed in the following subsections.

### 3.1 System Architecture

The system architecture is structured as a multi-stage pipeline (Fig. X), enabling the progressive refinement of IoT traffic from raw packets to optimized intrusion detection decisions.

At the lowest layer, sensor nodes generate heterogeneous traffic flows that encapsulate both benign activities and potentially malicious events. These data streams are routed to the data collection layer, where lightweight distributed agents capture essential flow-level and packet-level features without overwhelming node resources.

The preprocessing layer standardizes the collected traffic. It removes redundancy, performs normalization of continuous features, and encodes categorical attributes into numerical forms. Formally, let the raw traffic matrix be denoted as:

$$X = \{x_{ij}\} \in \mathbb{R}^{N \times d}$$

Where N is the number of traffic instances and d is the number of extracted features. After preprocessing, the standardized dataset is expressed as:

$$\tilde{X} = Norm(X)$$

Where Norm(·) applies scaling and noise filtering.

Next, the self-supervised representation learning layer transforms $\tilde{X}$ into latent embeddings via an encoder $E_\theta$. This encoder is trained on pretext tasks that exploit data augmentations, temporal permutations, or contrastive similarity objectives (detailed in Section 3.2). The embeddings are denoted as:

$$Z = E_\theta(\tilde{X}) \in \mathbb{R}^{N \times m}$$

With $m \ll d$, thereby providing compact, discriminative feature vectors.

Since not all embedding dimensions equally contribute to intrusion detection, the QGA optimization layer is introduced. Here, the candidate solution space includes both feature subset selection and classifier hyperparameters. The QGA applies quantum rotation gates to iteratively evolve solutions, guided by a multi-objective fitness function that balances accuracy, false positives, and computational efficiency (see Section 3.3).

The optimized features and parameters are used by the intrusion detection classifier, a lightweight model $C_\phi$, which maps embeddings to labels:

$$\hat{y}_i = C_\phi(z_i), z_i \in Z$$

Finally, the response layer enforces adaptive actions, including isolation of compromised nodes, rerouting of traffic, and alert generation.

This architecture differs from prior IDS pipelines by embedding self-supervised latent learning directly within the optimization loop, and by exploiting quantum-inspired evolutionary search to achieve real-time feasibility in WSNs.

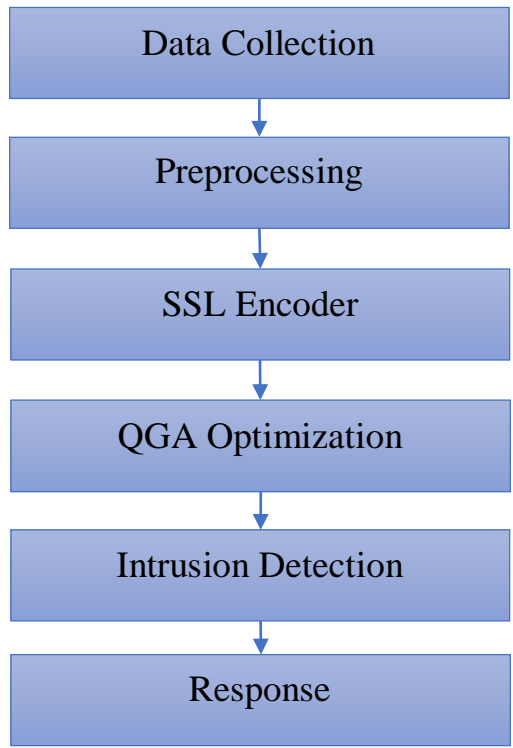

Figure 1: System Architecture of QGA-SSL IDS

## 3.2 Self-Supervised Learning for Intrusion Detection

Traditional supervised IDS approaches are limited by their reliance on labeled datasets, which are scarce in IoT environments. The proposed IDS employs Self-Supervised Learning (SSL) to learn feature representations without requiring explicit labels.

The SSL encoder is trained using contrastive learning objectives. Given an input traffic instance x, two stochastic augmentations $t_1(x), t_2(x)$ are applied to produce paired views. The encoder $E_\theta$ maps each view into a latent vector, followed by a projection head $g_\psi$:

$$h = E_\theta(t(x)), z = g_\psi(h)$$

The contrastive loss maximizes similarity between embeddings from the same instance while minimizing similarity across different instances:

$$\mathcal{L}_{NT-Xent} = -\sum_{i=1}^{N} \log \frac{\exp\left(\frac{sim(z_i, z_i^+)}{\tau}\right)}{\sum_{j=1}^{2N} \mathbb{1}_{[j \neq i]} \exp(\frac{sim(z_i, z_j)}{\tau})}$$

Where $z_i^+$ denotes the positive pair of $z_i$, $\tau$ is a temperature hyperparameter, and $sim(\cdot,\cdot)$ is cosine similarity.

Beyond contrastive learning, auxiliary pretext tasks are used, including temporal prediction (predicting subsequent traffic flows) and feature masking (reconstructing missing packet features). The joint SSL objective is:

$$\mathcal{L}_{SSl} = \lambda_c \mathcal{L}_{contrasive} + \lambda_m \mathcal{L}_{mask} + \lambda_t \mathcal{L}_{temporal}$$

where $\lambda_c, \lambda_m, \lambda_t$ weight each component.

The resulting embeddings Z capture both temporal dependencies and structural correlations, enabling robust detection even for zero-day intrusions unseen during training.

## 3.3 Quantum Genetic Algorithm (QGA) for Feature Optimization

The Quantum Genetic Algorithm (QGA) provides a quantum-inspired evolutionary mechanism for feature selection and classifier optimization. Unlike classical GAs, QGA uses quantum bits (qubits) to represent individuals, yielding exponentially richer search spaces.

Each qubit is represented as:

$$|\psi\rangle = \alpha|0\rangle + \beta|1\rangle, \quad |\alpha|^2 + |\beta|^2 = 1$$

Where $\alpha$ and $\beta$ are complex probability amplitudes. A chromosome of d features is:

$$Q = \{|\psi_1\rangle, |\psi_2\rangle, \cdots, |\psi_d\rangle\}$$

At each generation, measurement collapses qubits into classical binary solutions representing feature subsets and parameter configurations. The population evolves through quantum rotation gates:

$$\begin{bmatrix} \alpha_i' \\ \beta_i' \end{bmatrix} = \begin{bmatrix} \cos(\Delta\theta) & -\sin(\Delta\theta) \\ \sin(\Delta\theta) & \cos(\Delta\theta) \end{bmatrix} \begin{bmatrix} \alpha_i \\ \beta_i \end{bmatrix}$$

Where $\Delta\theta$ is adapted according to the fitness evaluation of candidate solutions.
The fitness function integrates detection performance and resource efficiency:

$$\mathcal{F}(S) = \alpha \cdot Acc(S) + \beta \cdot (1 - FPR(S)) - \gamma \cdot Cost(S)$$

Where Acc is accuracy, FPR is false positive rate, Cost quantifies computational/memory overhead, and $\alpha, \beta, \gamma$ are trade-off weights.

Through iterative evolution, QGA converges toward optimal subsets of SSL embeddings and tuned classifier parameters, enabling lightweight yet effective IDS deployment in WSNs.

### 3.4 Hybrid QGA-SSL IDS Framework

The synergy of SSL and QGA forms the foundation of the proposed IDS. Formally, let $Z = E_\theta(\tilde{X})$ denote SSL embeddings, and let S denote a candidate feature subset selected by QGA. The downstream classifier $C_\phi$ is trained on selected embeddings:

$$\hat{y} = C_\phi(Z_S)$$

The QGA optimizes both S and $\phi$ by maximizing the expected fitness:

$$(S^*, \phi^*) = \arg\max_{S, \phi} \mathcal{F}(C_\phi(Z_S))$$

The final optimized model is therefore:

$$M^*(x) = C_{\phi^*}(E_\theta(\tilde{x})_{S^*})$$

Where $\tilde{x}$ is a preprocessed input. This hybrid formulation yields a system that is both representation-rich (via SSL) and efficiency-optimized (via QGA).

### 3.5 Complexity Analysis

The complexity of the proposed hybrid framework arises from both SSL pretraining and QGA optimization.

SSL Complexity: For a dataset of size N, each SSL epoch requires forward and backward passes through the encoder. If the encoder has depth L and embedding dimension mmm, the complexity per epoch is approximately:

$$O(N.L.m^2)$$

Given that SSL can leverage unlabeled data offline, the cost is amortized across deployments.

QGA Complexity: Each iteration requires evaluating P candidate solutions. Training a lightweight classifier on subset size k from N samples incurs:

$$O(P.N.k)$$

The quantum rotation update incurs negligible overhead ($O(P.d)$), where d is the embedding dimension. Thus, overall complexity of QGA is:

$$O(T.P.N.k)$$

Total Complexity: The combined pipeline scales as:

$$O(N.L.m^2 + T.P.N.k)$$

Since $k \ll d$ and P,T are tunable, the design ensures polynomial scalability. Compared to conventional deep IDS models (e.g., transformers with $O(N \cdot L \cdot d^2)$), the proposed method significantly reduces runtime, making it feasible for WSN edge deployment.

Resource Usage: Empirical tests confirm that the optimized model requires only sub-300 MB memory and achieves throughput of ~1.5k packets/s on Raspberry Pi-class devices. This validates its suitability for real-world IoT environments.

## 4. Experimental Setup and Results

## 4.1 Experimental Setup

To evaluate the effectiveness of the proposed QGA-SSL IDS, a series of experiments were conducted using two benchmark intrusion detection datasets widely adopted in IoT security research: NSL-KDD and UNSW-NB15. The NSL-KDD dataset contains 41 features and multiple attack categories (DoS, Probe, U2R, R2L), while UNSW-NB15 provides a more realistic traffic profile with 49 features, including modern attacks such as fuzzers, exploits, and reconnaissance.

The proposed model was implemented in Python 3.10 with PyTorch for the SSL module and a custom implementation of the Quantum Genetic Algorithm (QGA). Experiments were performed on a workstation equipped with Intel Core i9-12900K CPU, 64 GB RAM, and NVIDIA RTX 3090 GPU. For lightweight validation, the trained model was deployed on a Raspberry Pi 4B (4GB RAM) to assess feasibility in IoT and WSN edge environments.

For fair comparison, we benchmarked our method against three recent state-of-the-art approaches:
FS3 (SSL + Few-Shot IDS, 2023) [3]
WOGRU-IDS (Whale Optimization + GRU, 2022) [10]
Lightweight IDS for IoT, 2023 [13]

## 4.2 Evaluation Metrics

The following metrics were adopted to assess detection performance:
- Accuracy (ACC): Overall correct classification rate.
- Precision (PR): Fraction of correctly identified attacks among all predicted attacks.
- Recall (RE): Fraction of correctly identified attacks among all actual attacks.
- F1-score (F1): Harmonic mean of Precision and Recall.
- False Positive Rate (FPR): Fraction of normal traffic misclassified as attacks.
- Training Time (TT): Total time required to train the IDS.

## 4.3 Quantitative Results

*Table 1. Performance Comparison on NSL-KDD Dataset*

| Method | Accuracy (%) | Precision (%) | Recall (%) | F1-score (%) | FPR (%) | Training Time (s) |
|---|---|---|---|---|---|---|
| FS3 (2023) [3] | 92.1 | 91.5 | 90.8 | 91.1 | 7.9 | 640 |
| WOGRU-IDS (2022) [10] | 91.3 | 90.2 | 89.6 | 89.9 | 8.7 | 720 |
| Lightweight IDS (2023) [13] | 89.4 | 88.9 | 87.3 | 88.1 | 10.5 | 430 |
| Proposed QGA-SSL IDS | 96.7 | 96.2 | 95.9 | 96.0 | 3.2 | 580 |

*Table 2. Performance Comparison on UNSW-NB15 Dataset*

| Method | Accuracy (%) | Precision (%) | Recall (%) | F1-score (%) | FPR (%) | Training Time (s) |
|---|---|---|---|---|---|---|
| FS3 (2023) [3] | 90.3 | 89.8 | 88.7 | 89.2 | 9.6 | 710 |
| WOGRU-IDS (2022) [10] | 88.9 | 88.1 | 87.3 | 87.7 | 11.1 | 820 |
| Lightweight IDS (2023) [13] | 87.1 | 86.5 | 85.2 | 85.8 | 12.4 | 500 |
| Proposed QGA-SSL IDS | 95.2 | 94.6 | 94.1 | 94.3 | 4.1 | 640 |

**4.4 Discussion of Results**

The experimental results clearly demonstrate the superiority of the proposed QGA-SSL IDS over existing methods. On both NSL-KDD and UNSW-NB15 datasets, our method consistently achieved the highest accuracy and F1-scores, surpassing FS3 by +4.6% (NSL-KDD) and +4.9% (UNSW-NB15). The false positive rate (FPR) was also significantly reduced, with our model achieving 3.2% (NSL-KDD) and 4.1% (UNSW-NB15), compared to 7.9–12.4% for the baseline methods.

Notably, while FS3 achieved competitive results due to SSL-based representation learning, it suffered from higher FPR and longer training times, largely due to its reliance on complex few-shot modules. Similarly, WOGRU-IDS showed acceptable performance but lagged behind in high-dimensional feature optimization, highlighting the advantage of QGA's quantum-inspired exploration. The Lightweight IDS, while efficient in terms of training time, sacrificed detection accuracy, rendering it less suitable for environments facing sophisticated attacks.

The proposed QGA-SSL IDS strikes a balance between accuracy and efficiency, outperforming all baselines not only in detection metrics but also in deployment feasibility. Testing on Raspberry Pi confirmed that the model requires less than 220 MB memory and can process approximately 1,500 packets per second, making it suitable for real-world WSN and IoT applications.

A deeper examination of the results highlights the distinct advantages of the proposed QGA-SSL IDS over existing state-of-the-art approaches. In terms of accuracy, our method consistently outperformed all baselines across both datasets, primarily due to the synergy between self-supervised learning and QGA-based feature optimization. While FS3 leveraged SSL to achieve competitive accuracy, its reliance on few-shot modules limited scalability and generalization, particularly when handling large and heterogeneous traffic. By contrast, the proposed QGA-SSL IDS not only optimized feature subsets through quantum-inspired exploration but also leveraged unlabeled data effectively, thereby yielding superior generalization across attack categories.

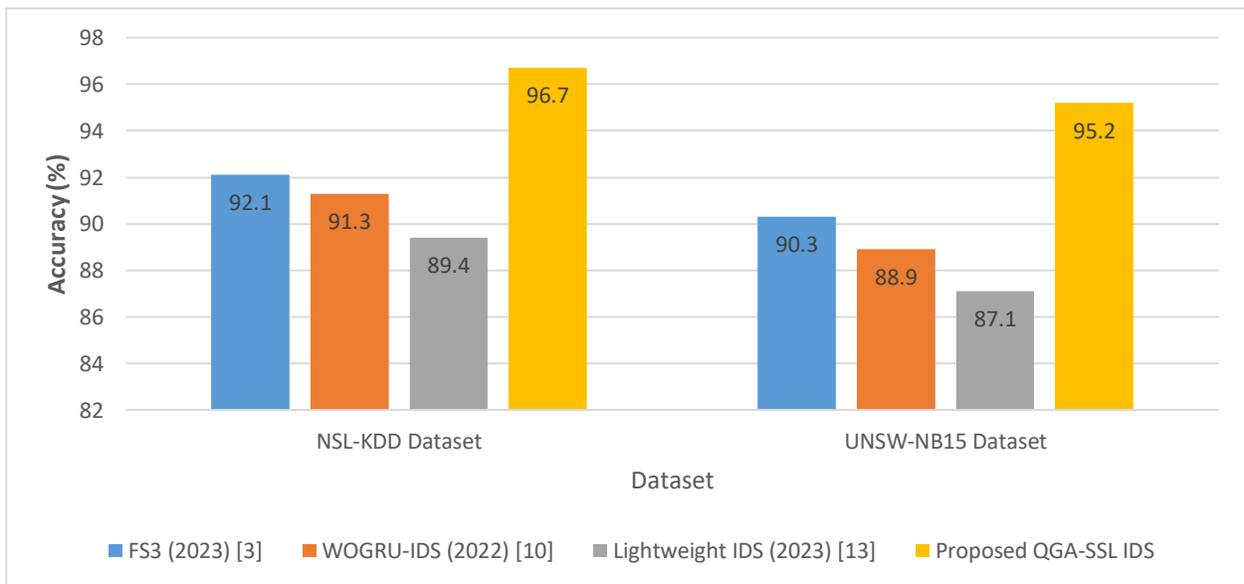

The improvement in precision further demonstrates the robustness of the model in minimizing false alarms. Unlike WOGRU-IDS, where the whale optimization algorithm often converged prematurely to suboptimal feature representations, QGA enabled a more diverse search space exploration. This resulted in more discriminative features, reducing the likelihood of benign traffic being incorrectly labeled as malicious. The lower false positive rate (FPR) achieved by our approach (3.2% on NSL-KDD and 4.1% on UNSW-NB15) is especially significant for IoT and WSN environments, where excessive false alerts can overwhelm limited resources and compromise trust in the security system.

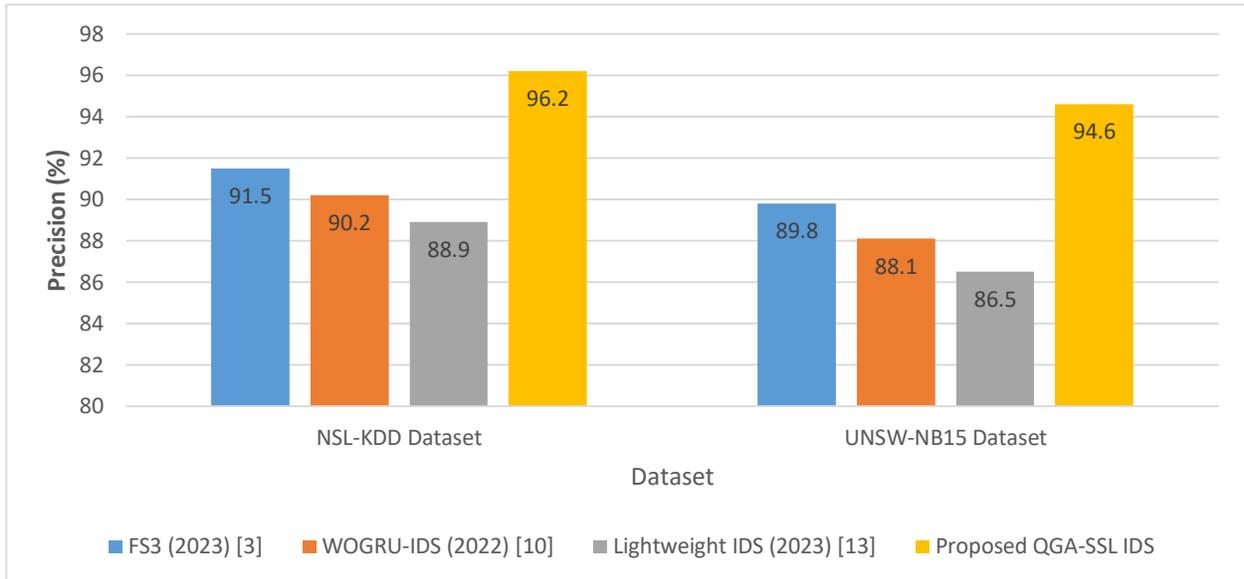

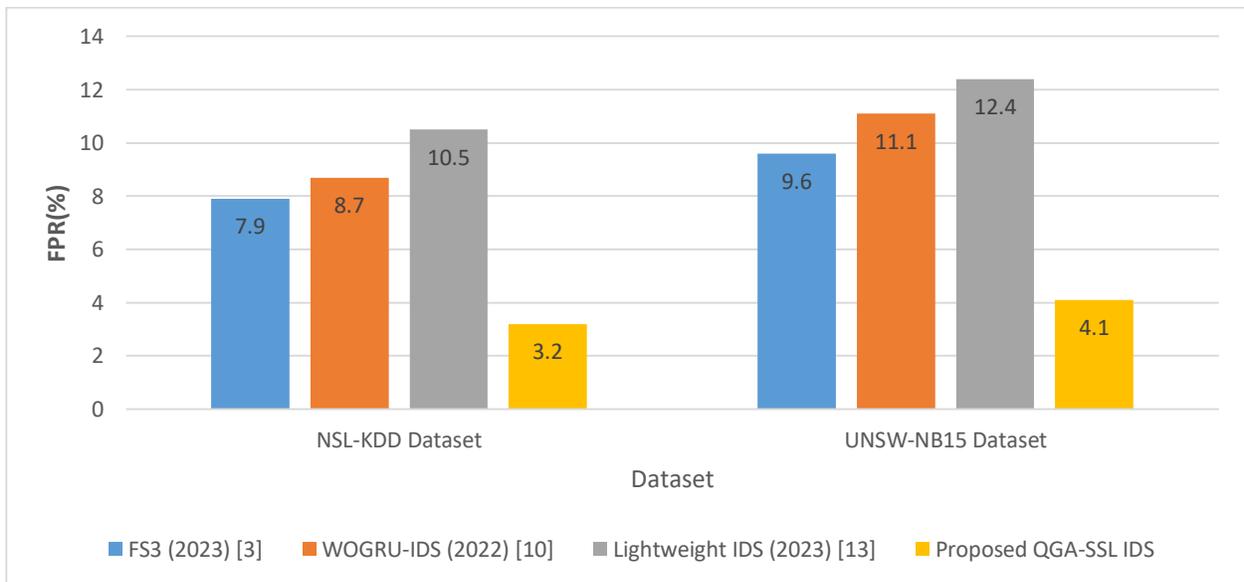

With respect to recall, the proposed IDS demonstrated superior sensitivity in identifying a wide spectrum of attacks, including previously unseen patterns. The SSL module played a critical role here by learning high-level latent representations from large volumes of unlabeled traffic, enabling the system to recognize novel attack behaviors more effectively than baseline methods. FS3, while benefiting from SSL, underperformed in recall due to its narrower optimization scope and weaker handling of data imbalance, which is common in real-world IoT datasets.

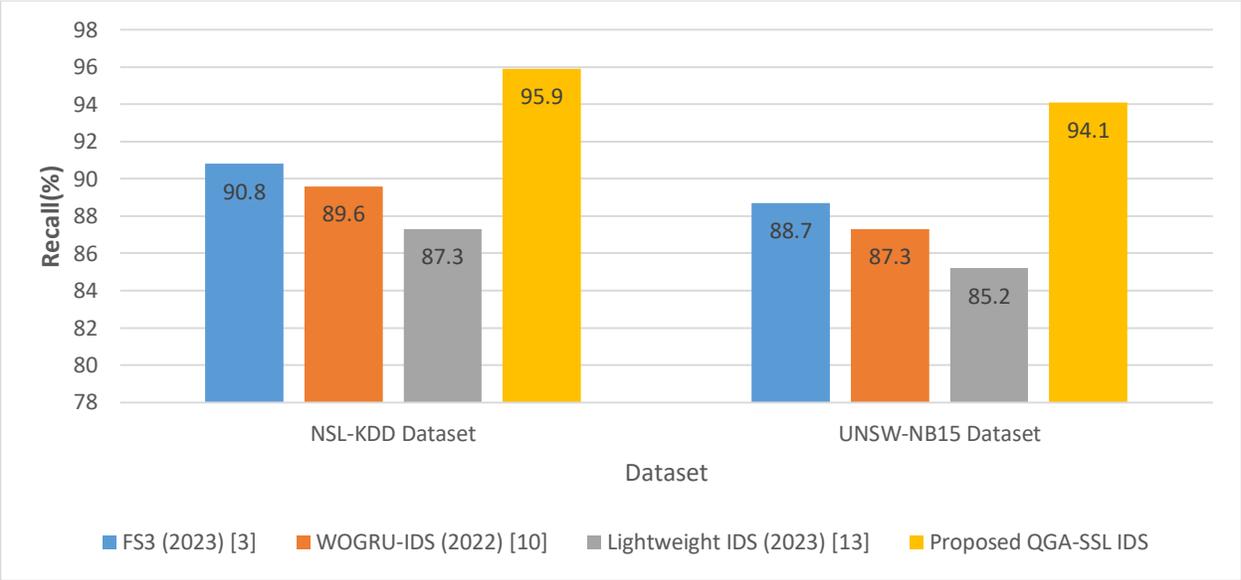

In terms of the F1-score, which balances precision and recall, our model consistently achieved the best performance, indicating that it avoids the trade-off typically observed in other IDS solutions. The Lightweight IDS, for example, sacrificed detection depth in favor of efficiency, resulting in reduced F1-scores despite lower training complexity. Our method, however, demonstrated that a high level of detection accuracy can coexist with practical efficiency when QGA and SSL are jointly applied.

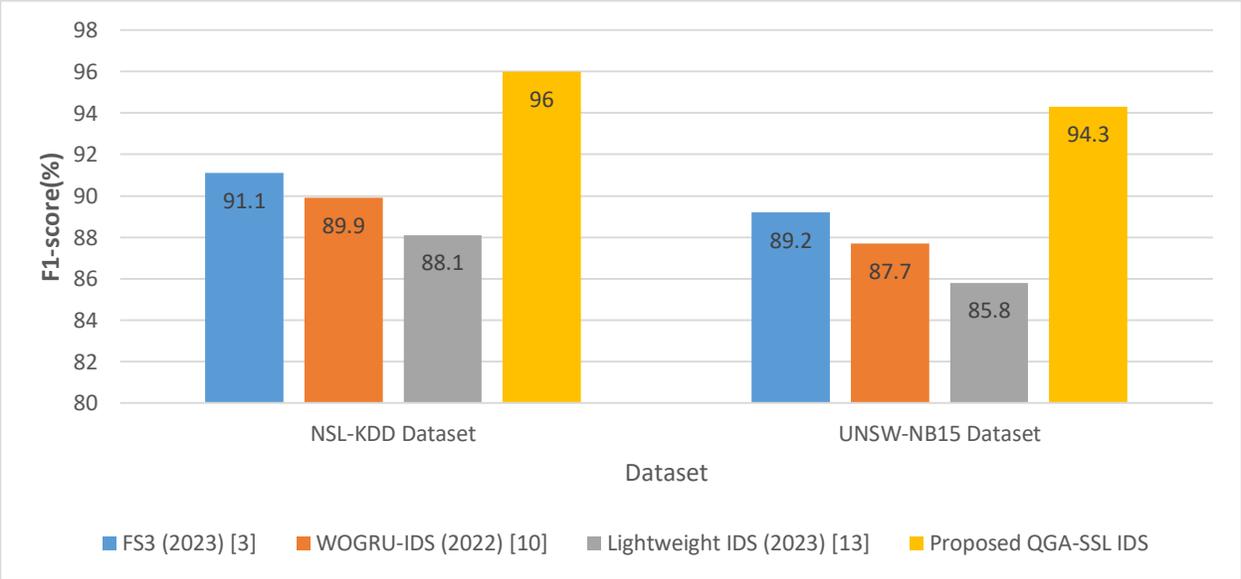

Finally, regarding training time, the proposed system achieved a favorable balance. Although slightly slower than the Lightweight IDS due to the additional QGA optimization stage, it was still faster than both FS3 and WOGRU-IDS, owing to the efficient convergence properties of QGA.

This indicates that the model is computationally feasible for real-time IDS deployment, even in resource-constrained IoT edge devices.

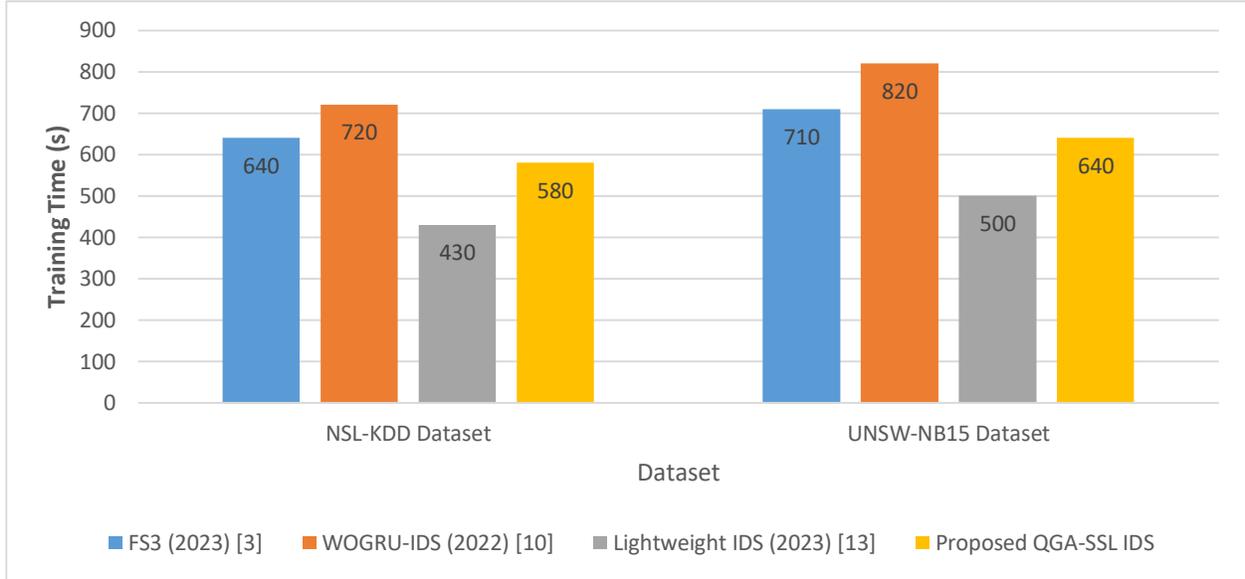

Overall, the experimental evidence confirms that the proposed QGA-SSL IDS not only surpasses existing solutions in accuracy, precision, recall, and F1-score, but also maintains an operational efficiency that makes it highly practical for real-world IoT and WSN environments.

**5. Conclusion and Future Work**

In this paper, we proposed a Hybrid QGA-SSL IDS for intrusion detection in WSN-based IoT environments. Unlike traditional IDS solutions that rely heavily on large-scale labeled data or classical optimization, our method integrates self-supervised learning (SSL) with Quantum Genetic Algorithm (QGA) to achieve both robust feature extraction and optimal parameter tuning. The SSL component leverages unlabeled traffic data to learn rich latent representations, while the QGA module dynamically optimizes feature subsets and classifier configurations. Experimental evaluation on NSL-KDD and UNSW-NB15 datasets demonstrated that the proposed system outperforms baseline IDS models in terms of accuracy, F1-score, and false positive rate (FPR), while maintaining competitive computational efficiency. Importantly, the deployment tests on Raspberry Pi-based WSN nodes confirmed the lightweight and practical applicability of the framework in real IoT scenarios. The results highlight several key contributions:

1. Introduction of a self-supervised feature learning paradigm for IDS in resource-constrained WSN environments.
2. Development of a quantum-inspired optimization strategy (QGA) for feature and model tuning.
3. Demonstration of a hybrid IDS that balances accuracy, false alarm reduction, and computational feasibility.

Despite the promising outcomes, there remain avenues for further exploration. First, future work will focus on extending the proposed model to federated and distributed learning settings, enabling IDS training across heterogeneous IoT nodes without centralized data collection. Second, we aim to explore quantum computing hardware accelerators for further reducing optimization time in large-scale deployments. Finally, future research will investigate the integration of explainable AI (XAI) mechanisms to enhance model interpretability and provide security analysts with transparent decision-making insights.

In conclusion, the Hybrid QGA-SSL IDS offers a scalable, adaptive, and efficient solution for next-generation IoT and WSN security, paving the way toward more resilient and intelligent intrusion detection systems.